\pdfoutput=1

\documentclass[sigconf,screen]{acmart}
\AtBeginDocument{%
  \providecommand\BibTeX{{%
    \normalfont B\kern-0.5em{\scshape i\kern-0.25em b}\kern-0.8em\TeX}}}

\setcopyright{acmlicensed}
\copyrightyear{2024}
\acmYear{2024}
\acmDOI{XXXXXXX.XXXXXXX}

\acmConference[FSE '24]{International Conference on the Foundations of Software Engineering}{Mon 15 - Fri 19 July 2024}{Porto de Galinhas, Brazil, Brazil}
\acmBooktitle{Proceedings of the 32nd ACM Symposium on the Foundations of Software Engineering (FSE '24), November 15--19, 2024, Porto de Galinhas, Brazil}
\acmISBN{978-1-4503-XXXX-X/18/06}

\usepackage{algorithm}  
\usepackage{algpseudocode}
\usepackage{threeparttable} 
\usepackage{listings}

\algnotext{EndIf}
\algnotext{EndFor}
\algnotext{EndFunction}

\newcounter{researchquestion}[section]
\renewcommand{\theresearchquestion}{\arabic{researchquestion}}

\newenvironment{researchquestion}[1]{%
  \refstepcounter{researchquestion}%
  \par\medskip\noindent
  \begin{minipage}{\dimexpr\linewidth-2\fboxsep-2\fboxrule}
  \textbf{Research Question \theresearchquestion:} #1
  \end{minipage}\par\medskip
}{}

\newcounter{researchquestionanswer}[section]
\renewcommand{\theresearchquestionanswer}{\arabic{researchquestionanswer}}

\newenvironment{researchquestionanswer}[1]{%
  \refstepcounter{researchquestionanswer}%
  \par\medskip\noindent
  \fbox{\begin{minipage}{\dimexpr\linewidth-2\fboxsep-2\fboxrule}
  \textbf{Answer to Research Question {\theresearchquestionanswer}:} #1
  \end{minipage}}\par\medskip
}{}

\begin{document}


\title{Observation-based unit test generation at Meta}


\author{Nadia Alshahwan}
\orcid{0009-0009-4763-0396}
\authornote{Author order is alphabetical. The corresponding author is Mark Harman. }
\author{Mark Harman}
\orcid{https://orcid.org/0000-0002-5864-4488}
\author{Alexandru Marginean}
\orcid{https://orcid.org/0009-0001-5311-762X}
\author{Rotem Tal}
\orcid{https://orcid.org/0009-0002-4320-070X}
\author{Eddy Wang}
\orcid{https://orcid.org/0009-0009-8825-6986}
\affiliation{%
  \institution{Meta Platforms Inc.,}
  \streetaddress{1 Hacker Way}
  \city{Menlo Park}
  \state{California}
  \country{USA}
}

\renewcommand{\shortauthors}{Alshahwan and Harman, et al.}

\begin{abstract}
TestGen automatically generates unit tests, carved from serialized observations of complex objects, observed during app execution.
We describe the development and deployment of TestGen at Meta. 
In particular, we focus on the scalability challenges overcome during development in order to deploy observation-based test carving at scale in industry.
So far, TestGen has  landed 518 tests into production, which have been executed 9,617,349 times in continuous integration, finding 5,702 faults. 
Meta is currently in the process of more widespread deployment.
Our evaluation reveals that, when carving its observations from 4,361 reliable end-to-end tests, TestGen was able to generate tests for at least 86\% of the classes covered by end-to-end tests. 
Testing on 16 Kotlin Instagram app-launch-blocking tasks demonstrated that the TestGen tests would have trapped 13 of these before they became launch blocking.       
\end{abstract}




\maketitle

\section{Introduction}

This paper describes our experience developing and  deploying observation-based TestGen (`TestGen-obs' or `TestGen'  for short\footnote{We also have an entirely separate tool, called `TestGen-LLM' \cite{mhetal:TestGen-LLM}, which {\em extends existing human-written} test cases using Assured LLM-Based Software Engineering \cite{mhetal:intense24-keynote}. By contrast, TestGen-obs generates new tests from scratch using observations.}).
TestGen is a tool that automatically generates unit and regression tests from scratch using scalable serialized observations of salient objects created during app executions; the tests are generated in Kotlin in order to  test  Meta's Java and Kotlin code bases.

Our intention with TestGen is not to fully replace human-written unit and integration testing.
Nevertheless, at the scale Meta operates (hundreds of millions of lines of code and over 3 billion users) it is infeasible to rely {\em solely} on human engineering effort for test case design.

Therefore, we have  embarked on an automation program for over a decade, using both well-established and novel software engineering research to deploy automated test design at scale.
This process started with the deployment of end-to-end automated test design tools such as Sapienz \cite{mhetal:ssbse18-keynote}, and static analysis tools such as Infer \cite{infer-open-source}.
We went on to deploy automated simulation-based test generation for higher level testing of whole communities of inter-acting users \cite{jaetal:mia,tuli:simulation,idmh:ase22-keynote}.

This previous work covered the system level testing requirements such as end-to-end and social testing. 
More recently, in order to shift testing leftwards in the development process \cite{alshahwan:software}, Meta has been targeting unit test design automation. 
Consequently, we are now tackling the problem of automatic generation of integration and unit tests, hence TestGen.\looseness=-1

A TestGen  test is a machine-generated unit test that captures the current behavior of the function under test, and thus it is capable of detecting regressions. 
The tool works by first automatically instrumenting the app to record runtime values of functions for which we want to generate unit tests.
It records current object instances, return values and parameter values. 
We call these recorded runtime values `observations'. 
TestGen automatically saves the observations when the instrumented version of the app executes the target function under test. 
This can be done either by running the app manually or using existing app execution tooling,
such as Jest
End-to-End (E2E) testing or the Sapienz test infrastructure, which has been deployed since 2018~\cite{mhetal:ssbse18-keynote,mao:sapienz:16}.\looseness=-1

A TestGen test asserts that, when called on the observed object instance with the observed parameter(s), the function under test produces the expected previously-observed return value. 
For the case of functions without a {\tt return}, TestGen tests  assert that the function executes without exception.

TestGen was originally implemented to generate unit tests for Meta's  Instagram app, so the results we report in this paper are based on TestGen's deployment for Instagram. 
However, there is nothing specific to Instagram in how TestGen operates and it is now being deployed across other Meta platforms and products.

TestGen tests are regression tests: they assert that the method under test in a diff\footnote{At Meta, a pull request is known as a `diff' (short for `differential'), following the Mercurial repository management nomenclature.} 
behaves the same as it did when executed on the current main branch (called `master'). 
This has 2 primary benefits:

\begin{enumerate}
\item 
It fully automates unit test generation, without the need for human intervention to define the assertions (there is no need for an oracle \cite{ebetal:oracle}).
\item 
It produces realistic test cases, thereby avoiding false positives. 
The tested values can occur at runtime because they have been observed in a previous execution of the app under exactly the same circumstances as those tested.\looseness=-1
\end{enumerate}

The primary contributions of this paper are:

\begin{enumerate}
\item A scalable industrial-strength observation-based unit test generation system: TestGen. 

\item 
A description of the principal novel technical features required to achieve sufficient scalability that allowed us to deploy observation-based testing at Meta: depth-aware serialization/deserialization, and an observation–aware Android memory manager.

\item A report of our experience of TestGen deployment, where it has thus far revealed 5,702 faults, and demonstrated its potential to trap at least 81\% of the important issues that might otherwise impact the launch of new versions of the Instagram app. 
\end{enumerate}

\section{Meta's Observation-Based TestGen System}

\begin{figure}[ht]
\includegraphics[width=80mm]{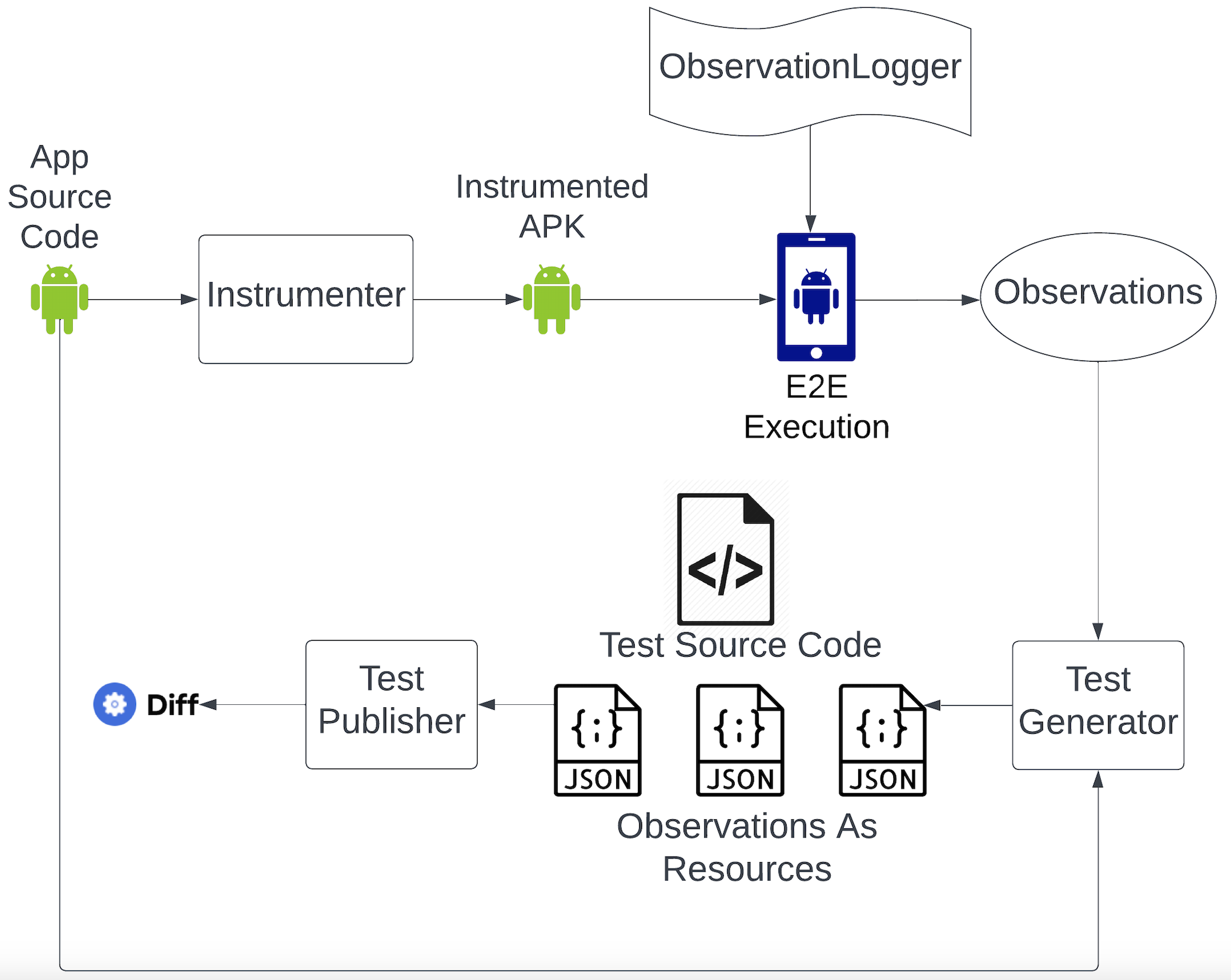}
\caption{The Architecture  of TestGen. TestGen instruments the app source code to record observations at runtime using the ObservationLogger. The TestGenerator uses the observations and the app source code to produce fully-runnable unit tests. Finally, the TestPublisher integrates with the build system and publishes a diff with the tests.   }
\label{fig:archi}
\end{figure}

Given a method under test, {\tt foo}, on an object of class {\tt A}, 
TestGen produces (potentially many) unit tests, 
each of which reflects a single execution trace  and asserts that:
When called on an observed object instance of type {\tt A}, 
with the parameters that TestGen has observed in a previous execution, {\tt foo} returns the same value as that observed in this previous app execution.
This is a form of Test Carving \cite{elbaum:carving}, as we explain in more detail in the 
Related Work section (Section~\ref{sec:related}).

TestGen designs its tests, based on observations from previous app executions. 
It collects these observations by building a specially instrumented version of the app. 
The instrumented app deploys unobtrusive and fast probes that collect the serialized values  subsequently used to build tests.  
The instrumentation logic does not affect the functional behavior of the app.
It simply collects observations silently in the background.  It is also designed to be lightweight, so as not to unduly affect app execution time, which might otherwise influence test behavior.

The instrumented app dumps observations into a data store.
Currently, we use the standard Android app DB as the data store that we optimized for a highly concurrent environment with a high volume of writes. Because the executions are real E2E executions of the whole app, the values observed are highly realistic and faithfully reflect those that could occur in production.


\autoref{fig:archi} depicts the architecture  of TestGen. 
The source code of the app is compiled with the `Instrumenter' compiler plugin that produces the `Instrumented APK'. At runtime, the `Instrumented APK' calls the `ObservationLogger'  to record  observations. 
The observations together with the initial app source code are compiled with the `Test Generator' plugin that constructs the tests, using the deserialized json representation of the observations. 
Finally, the `Test Publisher' constructs the build system files and dependencies for the  test cases, running them 5 times to confirm they are not flaky. 
For all tests that pass and are not flaky the TestPublisher produces a source control revision (i.e., a `diff') that can be landed. 
Once this happens, Meta's Continuous Integration (CI) platform automatically runs these tests, both continuously and also every time an engineer submits  a diff.

\noindent
\textbf{Test Maintenance:} Like any other Android unit test deployed at Meta, TestGen tests will block diffs from landing when they fail. 
It is thus important that signals that TestGen tests  provide are highly  actionable and that test maintenance does not become a burden for Meta engineers. 
It is also important  that TestGen provides automatic ways to unblock engineers when a test fails. TestGen relies on the assertion failure message to provide actionable information on how to unblock when a test fails. 
The assertion failure  exactly describes the difference that caused the failure:

\begin{lstlisting}[language=Java,breaklines=true, breakatwhitespace=true, postbreak=\mbox{$\hookrightarrow$},basicstyle=\scriptsize]
This is an automatically generated unit test from TestGen.
TestGen equality assertion failed! When comparing the return from 
the method under test with the expectation for the field `X` of 
the class `C`  we saw the value `true` while we expected the 
value `false`. The expected return for this test is in the 
resource file `path/to/resource` at line: x column: y.
If this change is not expected after the change in your diff, you 
can debug this test with the debugger in Android Studio
\end{lstlisting}
From this point, the  diff author has the following options:
\begin{itemize}
    \item If the test failure is not expected after the change, the diff author will debug their code to fix the issue.
    \item If the test failure is the result of an expected change in behavior then TestGen provides the following automated single-click resolutions:
    \begin{itemize}
        \item Update the assertion to reflect the new behavior.
        \item Call TestGen again to regenerate the test (if the change is more fundamental than a simple assertion failure).
        \item Delete the test if the diff author deems it to be no longer needed.
    \end{itemize}
\end{itemize}

The rest of this section describes  the implementation of the main TestGen components.

\subsection{Instrumenter Plugin}
\label{sec:instrumenter_plugin}

The Instrumenters are distinct compiler plugins that we implemented for Java and Kotlin. 
In order to collect observations at runtime, an `Instrumenter' adds extra code into the app at compile time, to explicitly log the observations. 


The Kotlin and Java compilers use an Intermediate Representation (IR). 
The TestGen Instrumenter plugin modifies the IR of the Kotlin/Java code rather than the source code.  
The compilers generate standard bytecode from their IR representations.
The compiler-generated bytecode thereby includes the instrumentation logic.
The Instrumented bytecode makes  calls to the `ObservationLogger' (discussed in \autoref{sec:obs_logger}) to do the heavy lifting of actually logging the observations. 
\autoref{fig:instrumentations} shows an example of this IR instrumentation added for a Kotlin function.

The Instrumenter constructs an instance of the \textit{ObservationLogger} at line 2 in \autoref{fig:instrumentations}, 
the first parameter of which  specifies that observations should be saved to the app DB (`APP\_DB'). 
The second parameter specifies that a random seed should not be used in this instance\footnote{For reproducability of the runs, TestGen also allows the use of a predefined random seed.}.
The third parameter specifies that the observations should be logged on a different thread (to improve scalability). 

The Instrumenter also adds instrumentation code that, when executed, will record observations for the:

\begin{itemize}
\item
\textbf{Current Object Instance}: The instrumenter adds a call to \lstinline+logObs+ 
at the start  of the method body, to log 
the value of the current object instance. 
Line 3 in \autoref{fig:instrumentations} shows an example.
 
\item
\textbf{Method Parameters}: The instrumenter adds calls to \lstinline+logObs+ after the logging call for current object instance.
It will add a call for each function parameter in turn. Lines 4--5 in  \autoref{fig:instrumentations} show two such examples. 

\item
\textbf{Method Returns}: At each {\tt return} statement, the Instrumenter adds a call to the \lstinline+logObs+ as a meaning–preserving transformation.
It creates a temporary variable, in which it holds the returned value, then passes this to \lstinline+logObs+, and finally returns it. Lines  8--17 in  \autoref{fig:instrumentations} show an example of this. 
The block that appears between lines 8--15 is the way in which TestGen is able to log returns with the semantics-preserving transformation.
\end{itemize}

\begin{figure*}
\small
\begin{lstlisting}[language=Java,numbers=left,breaklines=true, breakatwhitespace=true, postbreak=\mbox{$\hookrightarrow$},escapeinside={(*@}{@*)}]
fun <elided method name>(session: Session?, intent: Intent?): Boolean {
  (*@\color{gray} val tmp0\_obs: ObservationLogger = ObservationLogger(logTo = "APP\_DB", randomSeed = null, logOnDifferentThread = true)@*)
  (*@\color{gray}tmp0\_obs.logObs</* null */>(observedValue = <this>, lNr = -1L, colNr = -1L, fileName = "file/path/Class.kt", obsType = "CURRENT\_OBJ\_INST")@*)
  (*@\color{gray}tmp0\_obs.logObs</* null */>(observedValue = session, lNr = 142L, colNr = 31L, fileName = "file/path/Class.kt", obsType = "PARAMETER")@*)
  (*@\color{gray}tmp0\_obs.logObs</* null */>(observedValue = intent, lNr = 142L, colNr = 54L, fileName = "file/path/Class.kt", obsType = "PARAMETER")@*)
  var intent: Intent? = intent
  ...
  { // BLOCK
    (*@\color{gray} val tmp0: Boolean = @*){ // BLOCK
      val tmp2_elvis_lhs: Boolean? = <this>.<get-isTestedCondition>()
      when {
        EQEQ(arg0 = tmp2_elvis_lhs, arg1 = null) -> false
        else -> tmp2_elvis_lhs
      }
    }
    (*@\color{gray}tmp0\_obs.logObs</* null */>(observedValue = tmp0, lNr = 153L, colNr = 4L, fileName = "/file/path/Class.kt", obsType = "RETURN")@*)
    return tmp0
  }
}
\end{lstlisting}
\vspace{-1em}
\caption{Kotlin Intermediate Representation (IR) of an example instrumentation. \lstinline+tmp0_obs+ is an instance of the observation logger. \lstinline+logObs+ is its method that logs the observations to the app DB. 
The grayed code is code that the instrumenter has added.\looseness=-1}
\label{fig:instrumentations}
\end{figure*}

Although it is possible to instrument the whole app, it can be very time consuming to completely rebuild a whole app, such as Instagram \cite{mh:apr22-keynote}. 
As a result, TestGen limits the observation logging to the set of files for which tests must be generated.

\subsection{ObservationLogger}
\label{sec:obs_logger}
To tackle the observation logging scalability challenge, we introduce DASAD (Depth-Aware Serialization And Deserialization).
DASAD introduces  depth-aware serialization/deserialization together with  a pointer-based observation  sharing mechanism that uses a pointer-aware Android memory management layer.
This scales the collection of observations, their serialization and serialization protest generation.
With DASAD observations can be made from the app, while running, without and affecting its performance.

By contrast, a simple reflection-based approach to fully serializing each relevant object used by an app  would not scale: the instrumented app would soon stop responding (i.e., AnR\footnote{\url{https://developer.android.com/topic/performance/vitals/anr}}) since such a simple approach would end up serializing large amounts of entire Android heap store. 
Furthermore, the space required would also make this infeasible.
For example, at the scale of the Instagram App, which is characteristic of the larger of the  commercially available apps,  it 
takes hours to serialize a single object instance without DASAD, which would
clearly impact the performance of the app under test, rendering any attempt at observation-based test generation approach  infeasible.

In the remainder of this section, we explain how DASAD addresses the two scalability challenges: 

\begin{enumerate}
\item Serializing large complex objects,  
\item Supporting concurrent runtime environments.
\end{enumerate}

\subsubsection{Serializing Large And Complex Objects}

In order to serialize large objects, DASAD uses a combination of depth awareness and a pointer-based observation representation approach, as explained below.

\noindent
{\bf Depth-Aware Serialization: } 
Fortunately, for the TestGen use case, it is usually {\em unnecessary} to record the entire object for the purpose of generating tests. 
Instead,  a test case may only need to retrieve values observed in nested objects up to a given depth. 
This may also generalize to many other use cases, where scalability can be achieved in a similar depth-aware manner, but here we focus exclusively on the TestGen use case.
To illustrate how depth awareness can help tackle scale for the test generation use case, consider the following Java classes:

\begin{lstlisting}[language=Java,numbers=left,breaklines=true, breakatwhitespace=true, postbreak=\mbox{$\hookrightarrow$},basicstyle=\scriptsize]
class Foo{A a;}
class A {B b;}
class B {C c;}
\end{lstlisting}

Suppose we want to serialize an instance of {\tt Foo}. 
A full serialization will recursively contain all 3 fields: {\tt a}, {\tt b}, and {\tt c}. 
If we serialize with a depth limit of 1, then the serialization will contain only the field {\tt a}, in which field {\tt b} will be {\tt null}.
Depth-aware  serialization simply replaces all fields that are nested deeper than the given depth with {\tt null}.

\noindent
{\bf Pointer-Aware Serialization in a Android Memory Manager: } 
Even with the depth-aware serialization, the number of objects produced is too numerous and large for it to be feasible to record them all. 
The core intuition that we used to tackle this issue is that most of the objects are sharable by an E2E run (consider, for example, the Android App context, used in many functions with an identical instance).
To avoid  duplication we constructed a memory management system on top of Android that the serialization  uses to determine when objects are the identical, and can thus be shared in memory. 

Algorithm~\autoref{alg:serialization}  describes DASAD serialisation. 
The serialization-function \textit{serialize\_object} receives as input an object to be serialized. 
The variable \textit{isFirstSerialization} keeps track whether or not we are serializing the current object for the first time: line 17 gets the object id and checks whether it is in \textit{seenObjects}, a hash set of seen objects ids. 
The object id is a hash over the object state, such that it uniquely identifies an object. 
If it is the first serialization, lines 18 and 19 set the value of  \textit{isFirstSerialization} to true and add it into the `seenObject'. 
We do this in a syncronized block since this code will be called in a  highly concurrent environment. 
If the object was previously serialized, TestGen simply marks its serialization as \textit{POINTS\_TO\_OBJECT\_ID} and  stops the serialization process. 

At deserialization time, the algorithm looks for a full serialization for the same object id to reconstruct it. If the object is encountered for the first time, we proceed with each serialization by recursively calling the \textit{serialize\_object} function on each of its fields. The variable \textit{objSerialization} is a string that contains the json representation of the serialized object\looseness=-1. 


\begin{algorithm}
\scriptsize
\caption{\textbf{Pointer-Aware Object Serialization:} the algorithm that TestGen uses to log observations at scale: it keeps track of already serialized objects to avoid multiple serializations of the same objects. 
\looseness=-1}
\label{alg:serialization}
\begin{algorithmic}[1]

\Function{is\_primitive}{$object$}
\State \Comment{This function return true if the object is primitive and false otherwise.}
\EndFunction

\State

\Function{primitive\_serialization}{$object$}
\State \Comment{This function dumps the  representation of a primitive object.}
\EndFunction

\State

\Function{append\_field\_serialization}{$objSerialization$, $fName$,$fSerialization$}
\State \Comment{This function appends to the json representation of the object  (\textit{objSerialization}) with the serialization of the field named \textit{fName}: \textit{fSerialization}.}
\EndFunction

\State

\State $seenObjects \gets \text{HashSet<String>}$
\State

\Function{serialize\_object}{$object$}
    \If{is\_primitive($object$)}
    \State
        \Return primitive\_serialization($object$)
    \EndIf
    \State $isFirstSerialization \gets \text{false}$
    \State \text{synchronized} \{
        \If{$object$.get\_id() $\notin$ $seenObjects$}
            \State $isFirstSerialization \gets \text{true}$
            \State $seenObjects$.add($object$.get\_id())
        \EndIf
    \State \}
    \If{ !$isFirstSerialization$}
        \State
        \Return "POINTS\_TO\_" + $object$.get\_id()
    \EndIf
    \State $objSerialization \gets ""$
    \For{each $field$ in $object$.get\_fields()}
        \State $objSerialization$ $\gets$ append\_field\_serialization ($objSerialization$, $field$.name(), serialize\_object($field$))
    \EndFor
    \State
    \Return $objSserialization$
\EndFunction
\end{algorithmic}
\end{algorithm}


TestGen  has to handle  pointers in deserialization. 
TestGen applies Algorithm~\autoref{alg:deserialization} to instantiate  pointers, prior to the test's execution. 
In Algorithm~\autoref{alg:deserialization}, the entry point is the function \textit{get\_full\_ serialization\_map} (line 41). 
This function is applied across all observations, for all tests generated from the same E2E run. 
It returns a set of all full initializations across all these observations (this is recursive, becuase  each field in the recursive traversal of an object can contain other objects that are either full serializations or pointer serializations). The first step, is to construct the set of \textit{pointerKeys} (lines 42--44) that contains all the keys of serializations that are pointers. 

Lines 45-47 construct the list of all full serializations that have at least one pointer serialization. 
Having this list will later allow us to replace all pointer serializations, such that we obtain observations containing only full serialization (as the test runtime requires). 
Because of the recursive nature of Algorithm~\autoref{alg:serialization}, observations might still contain pointer serializations for various fields in their recursive traversal. This is what the rest of the logic in \textit{get\_full\_serialization\_map} tackles.

Lines 48-49 call the method \textit{init\_serialization}. This function parses each \textit{fullSerialization} in turn and replaces all its pointer serializations (e.g., across its fields) with full serializations. 
The underlying data structure for \textit{fullSerialization} preserves the pointer relationships after replacement.
That is,  if we replace a field that is a pointer serialization P2 in a full serialization S1 with its corresponding full serialization S2, the field will keep pointing to the full serialization S2 rather than copying it. 
Therefore, the algorithm need only parse this list once: further initializations will include the previously processed full serializations. 

\textit{fullSerializations} might contain fields that are still stored as pointer serializations until lines 48-49 finish execution. 
The algorithm replaces all pointer serializations in \textit{init\_serialization} across all the observations and these replacements are also reflected in earlier replacements (when a particular observation might not be fully initialized).
Therefore, the algorithm  guarantees that, from line 50,  there will be no pointer serializations remaining. 

Additionally, \textit{init\_serialization} does not process deeper than  the \textit{MAX\_DEPTH} limit because  those values would be unneeded in any case.\looseness=-1 

At this point, the \textit{full\_serialization}s will contain  recursion (since data types can be recursive). 
Additionally, although we restrict observations to \textit{MAX\_DEPTH} with the depth-aware serialization, \textit{init\_serialization} might overshoot this depth, because of initializing the pointer serialization. 
For example, suppose the app has a nested class structure like this:  {\tt  A \{ b: B \{c: C \{d: D \{ ...\}\} \}\}}. 
Suppose we have an object, {\tt ObjA}, of  type {\tt A} that has pointers to nested elements. 
These nested elements may become instantiated, with the effect that the effective depth for {\tt ObjA} will be as deep as its nesting, irrespective of the maximum depth to which DASAD serializes.
In this way, \textit{MAX\_DEPTH}, is merely the maximum depth to which serialization will process in a single traversal, 
but not the maximum depth that will be achieved overall.

\textit{fullSerializations} is a dictionary with values that can be pointers (line 16 in \autoref{alg:deserialization} adds these pointers). 
Thus, any processing of these pointers will affect their values in all the \textit{fullSerializations} in which they appear. 
Since next we want to remove elements that bypass \textit{MAX\_DEPTH}, we first need to deep copy (in line 50). 
For example,  suppose a  full serialization, A, with pointer P is already at \textit{MAX\_DEPTH - 1}. 
In this case, we want only depth 1 in pointer P. 
But P can also be in another full serialization, B, where it is at depth 1. 
Here we need to keep P with \textit{MAX\_DEPTH - 1}. 
The deep copy allows us to do this by making each occurrence of P in \textit{fullSerializations} a distinct value, 
rather than then pointers to the same value.


The final stage of \textit{get\_full\_serialization\_map} removes recursion and keeps elements only up to \textit{MAX\_DEPTH} in the call to \textit{remove\_rec} at lines 51--52. \textit{remove\_rec} is a recursive function that passes each full serialization. When it reaches \textit{MAX\_DEPTH}, it simply removes the field values. To remove recursion, it keeps track (in \textit{seen}) of the objects encountered when parsing a top level full serialization. 
If the recursive traversal reaches an already seen object, we must be in a recursive case and thus replace the value with a `recursion' annotation. 
The outputs of \textit{get\_full\_serialization\_map} are then simply used as a string replacement over the set of all observations, to obtain the set of all fully initialized observations that the test cases can use at runtime.

\begin{algorithm}
\scriptsize
\caption{\textbf{Pointer-Aware Object Deserialization:}  the algorithm that instantiates observations prior to generating tests. The pointer-aware serialization leaves pointers in observations as an optimizations. Before the test uses the observations, they need to be fully-initialized such that the objects can be constructed at runtime.}
\label{alg:deserialization}
\begin{algorithmic}[1]

\Function{pointer\_serializations}{$observation$}
\State \Comment{This function returns all the keys of the pointer serialization.}
\EndFunction

\State

\Function{full\_serializations}{$observation$}
 \State\Comment{This function returns all the full serializations .}
\EndFunction

\State

\Function{init\_serialization}{$p$, $key$, $obs$, $initialized$, $seen$, $depth$}
    \If{ $depth > MAX\_DEPTH$}
        \Return
    \EndIf

    \If{is\_instance($obs$, $list$)}
        \For{$i$, $it$ in enumerate($obs$)}
            \State init\_serialization($obs, i, it, initialized, seen, depth+1$)
        \EndFor
    \EndIf

    \If{!is\_instance($obs$, $dict$)}
        \Return
    \EndIf

    \State $hash \gets $ compute\_hash\_key($obs$)

    \If{$hash \in seen\_keys$}
         \Return
    \EndIf

    \If{is\_pointer\_serialization($obs$)}
        \State $p[key] \gets initialized[hash]$ 
    \EndIf

    \State $seenKeys \gets seenKeys | hash$

    \For{$k$, $val$ $\in$ $obs.items$}
        \If{is\_instance($val$, $dict$)}
            \State init\_serialization($obs,k,val,initialized,seen,depth+1$)
        \EndIf

        \If{is\_instance($val$, $list$)}
            \For{$i$, $it$ $\in$ enumerate($val$)}
                \State init\_serialization($val,i,it,initialized,seen,depth+1$)
            \EndFor
        \EndIf
    \EndFor
\EndFunction

\State

\Function{remove\_rec}{$p$, $key$, $obs$, $initialized$, $seen$, $depth$}

    \If{ $depth > MAX\_DEPTH$}
        \If{is\_instance($obs$, $dict$)}
             $parent[key] \gets \{\}$
        \ElsIf{is\_instance($obs$, $list$)}
            $parent[key] \gets []$
        \EndIf
    \EndIf

    \If{is\_instance($obs$, $list$)}
        \For{$i$, $it$ in enumerate($obs$)}
            \State remove\_rec($obs,i,it,initialized,seen,depth+1$)
        \EndFor
    \EndIf

    \If{!is\_instance($obs$, $dict$)}
         \Return
    \EndIf

    \State $hashKey \gets $ compute\_hash\_key($obs$)

    \If{$hashKey \in seen$}
        \State $p[key] = annotate\_recursion(p[key])$

        \State \Return
    \EndIf

    \State $seenKeys \gets seen | hashKey$

    \For{$k$, $val$ $\in$ $obs.items$}
        \State remove\_rec($obs,k,val,initialized,seen,depth+1$)
    \EndFor
\EndFunction

\State

\Function{get\_full\_serialization\_map}{$observations$}
    \State $pointerKeys \gets set()$
    \For{$observation$ in $observations$}
        \State $pointerKeys$.update(pointer\_serializations($observation$))
    \EndFor

    \State $fullSerializations \gets \{\}$
    \For{$observation$ in $observations$}
        \State $fullSerializations$.update(full\_serialization ($observation$))
    \EndFor

    \For{$s$ in $fullSerializations$}
        \State init\_serialization($\emptyset,\emptyset,s,fullSerializations,set(),0$)
    \EndFor

    \State $fullSerializations \gets$ deep\_copy($fullSerializations$)

    \For{$s$ in $fullSerializations$}
        \State remove\_rec($\emptyset,\emptyset,s,fullSerializations,set(),0$)
    \EndFor

    \State \Return $fullSerializations$ 
\EndFunction

\end{algorithmic}
\end{algorithm}


\subsubsection{Support For A Very High Concurrent Runtime Environment}
\label{subsec:concurrent}
Many observations will be serialized and logged concurrently. 
This leads to a third scalability challenge for the `ObservationLogger': support for highly concurrent writes to an external storage medium. 
To write the observations to an external storage for later processing, we use the Android app DB, in which we store them as soon as the `ObservationLogger' serializes this. To tackle the scalability challenge of this highly concurrent writing, we:

\begin{enumerate}
    \item Execute all the serialization logic and the saving of the observations to the App DB on different background threads, that are not prioritized by the OS. 
    \item Optimise the Android app DB for multiple concurrent writes. 
    For this, we enabled write ahead logging\footnote{\url{https://www.sqlite.org/wal.html}} and set synchronous to off\footnote{\url{https://www.sqlite.org/pragma.html\#pragma\_synchronous}} 
\end{enumerate}

Once the observations are written to the app DB, the next steps of TestGen  read them from the DB and store them as resources files, in json format, together with the generated test case source code.\looseness=-1
The generated test case thus consists of a Kotlin unit test that deserializes the serialized versions of objects stored in the resources files, and uses these as the parameters to be passed to the function under test.
It makes assertions about the returned values from those calls. 

\subsection{Test Generator}
\label{sec:testgenerator}

The test generator reads previously stored observations, and generates tests for methods decorated with {\tt @GenerateTestCases}. 
The test generator is also implemented as a compiler plugin.

The TestGenerator workflow consists of three top level steps:

\noindent
{\bf 
Traverse code under test:} 
Traverse the compiler’s Intermediate Representation (IR) of the method under test, using compiler specific APIs 
(the specific API depends on whether the TestGenerator is traversing Java and Kotlin). 
In this traversal, a `store' is constructed. 
The store contains static information required to produce the test cases, such as containing class, containing package, method return type, and the imports that the class under test requires (since many of them might be used in the unit test)

\noindent
{\bf 
Compute information required to generate tests:} 
A TestDataGenerator component to TestGen computes test data for each of the store objects collected in the traversal. 
The TestDataGenerator has multiple implementations, thereby allowing for different test data generation strategies. 
Currently, we have implemented the observation-based test generation,
as described in this paper.
However, we purposely left this component generic, so that other strategies could be incorporated (see Section~\ref{sec:future}).

\noindent
{\bf 
Write out test cases:}
The TestWriter component  produces fully executable unit tests, written in Kotlin\footnote{It would be technically trivial to adapt TestGen to write the test cases it generates in other languages; the choice of Kotlin was a matter of convenience only.}. 
An example of a single test case generated by TestGen can be found in \autoref{fig:example_test}.
As can be seen, the format of the test is extremely simple, and human-readable, and the  code that expresses the test case uses standard unit testing formats.
All generated tests include a `kill switch', which allows us to globally switch  off all generated test cases in case of problems

\begin{figure}
\scriptsize
\begin{lstlisting}[language=Java,numbers=left,breaklines=true, breakatwhitespace=true, postbreak=\mbox{$\hookrightarrow$},basicstyle=\scriptsize]
class <elided test class name> {
  private val serializer: DASAD = 
    DASAD.getDASAD()
  private val TESTGEN_RUN_COMMAND =
    "cd ~<elided path name> && ./<elided script name>.sh" + 
    "$ARGS_USED_BY_TESTGEN_TO_GENERATE_THE_CURRENT_FILE"
  private val BUCK_INVOCATION_COMMAND =
    "buck test $BUCK_TARGET_GENERATED_FOR_THIS_TEST"

  @Test
  fun `test for $METHOD_UNDER_TEST_NAME $TEST_UUID`() {
    if (DASAD.TESTGEN_KILLSWITCHED) { return; }
    val subject: $CLASS_UNDER_TEST =
        DASAD.initializeObservation(
            "$PATH_TO_CURRENT_OBJECT_INSTANCE_OBSERVATION",
            javaClass,
            $CLASS_UNDER_TEST::class.java)
    val ret =
        subject.$METHOD_UNDER_TEST(
            DASAD.initializeObservation(
                "$PATH_TO_PARAMETER_1_OBSERVATION_IN_RESOURCES",
                javaClass,
               $CLASS_OF_PARAMETER_1::class.java),
	// ... the same pattern for all the parameters of the MUT
)
    assertTrue(
        DASAD.testGenAssertEqual(
            "$PATH_TO_RETURN_OBSERVATION",
            javaClass,
            ret,
            serializer,
            TESTGEN_RUN_COMMAND,
            BUCK_INVOCATION_COMMAND))
  }
}
// End of auto-generated text.
\end{lstlisting}
\vspace{-1em}
\caption{An Example  generated test class. Some commercially sensitive details have been elided. The test method: `test for \$METHOD\_UNDER\_TEST\_NAME \$TEST\_UUID` asserts that when called with the previously observed parameters on the previously observed object state, the method returns the previously observed value (lines 26--34)  }
\label{fig:example_test}
\end{figure}

Placeholders (starting with \$) are instantiated using static analysis.
The tests assert that, when the test is executed with respect to a previous observation of the same method, in same object state  (i.e., current object instance; lines 15--19), and with the same inputs (i.e., parameters; lines 20--27), then the method under test produces the same result (i.e., the same returned value) as the one previously observed (lines 28--36).

When the method under test has no return, the template changes slightly. 
It will only call the method under test, without adding an assertion. 
This makes the tests less valuable than the tests that assert something about a returned value, 
but they still bring value by asserting that the method under test does not crash; essentially falling back on the implicit oracle \cite{ebetal:oracle}.

\subsubsection {Assertion Equality}
\label{sec:testGenAssertEqual}
An important component of the generated tests is the equality assertion between the value returned by the method under test and the value stored as a previous  observation. 
Generally, TestGen  cannot assume that it is possible to assert equality between two arbitrary Android objects: this will only work when their class happens to override the equality operator (otherwise TestGen would be asserting pointer equality not value equality). 
To ensure the TestGen tests perform meaningful equality assertions, we implemented a custom equality assertion: {\tt testGenAssertEqual}.

{\tt testGenAssertEqual} takes a serialized observation and the runtime return of the method under test that should be equal in a passing test. 
It serializes to json the runtime return and compares the two json strings for equality. 
It considers only  common elements for equality (i.e., commonly existing fields in the recursive traversal of the data type), since fields that are initialized in a unit test may differ from those in the E2E run. 
That is,  in the E2E run, many methods other than the one under test will have been called. 
These methods can, for example, call the setter on a field that is not called in the unit test run. 
As long as the common fields are equal, and  test execution does not crash, then TestGen determines that the equality test has passed.\looseness=-1 



\subsection{Test Publisher}
\label{sec:selection}
The test publisher is the final high level TestGen component executed by an overall run of the TestGen workflow. 
It takes, as input, the test source file as produced by the Test Generator, and produces a diff that contains runnable reliable (i.e., not flaky) tests in CI (i.e., including the needed buck build files). 
To do this, it employs the following steps:

\noindent
{\bf 
Buck file:} 
A buck file is a form of `make' file, that describes how to build a particular component, called a ``buck target''.
The buck file generator automatically detects when (and where in the file system) tests have been generated, and automatically generates corresponding buck files at the right directory locations. 
The generated buck files implement the necessary dependencies, visibility rules and other buck configuration details required to ensure that the tests can be fully executed.\looseness=-1  

Tests can be executed from the command line using a {\tt buck test} command.
This is the normal way in which engineers execute unit tests on demand; TestGen tests are, after all, just normal tests, albeit machine-generated.
TestGen tests will also automatically be selected when a submitted diff contains code covered by the tests.\looseness=-1

\noindent
{\bf 
Buck Dependencies:} 
The challenge in automatically generating a buck target is to be sure that it contains all the required dependencies. 
The first category of dependencies includes those required by any  and all unit tests. 
Since the tests will directly call the methods on the class under test and use various types imported there (e.g., parameters and return types), we also automatically add all the dependencies of the class under test to the dependencies of the unit test's buck target. 

The most challenging dependencies to determine automatically are those required for observation deserialization. 
These can be arbitrary dependencies used in the recursive traversal of any data type used by the current object instance, returns or parameters. 
To compute them, we first analyze all the object types used across all observations. 
For each unique object, we construct a set of unique buck dependencies containing them. 
We then add all these dependencies to the buck dependencies of the test. 

In this way,  we ensure that all the required dependencies are included. 
The dependence computation process is `liberal'.
Therefore, it favors the generation of a correctly executable test, rather than one that might fail due to insufficient dependence information in the buck files. 
The buck file so-generated may  potentially include some extra dependencies (e.g., not all dependencies of the class under test are used in the unit test), but this is relatively harmless. 
The existing build system often catches and removes any unneeded dependencies using an independent workflow designed to cater for arbitrary buck files.

\noindent
{\bf 
Test Runner:} 
The Test Publisher will run all the tests in the newly-added buck target and will collect runtime information about them: whether the test passes consistently, coverage data, etc. 
The test runner deletes tests that are either broken or flaky so that the Test Publisher  does not include any tests that do not consistently pass or that are broken.
This insures that the signal to engineers from automatically-generated test is of the highest quality.

\noindent
{\bf 
Test Selection:}
Test generation can produce many test cases; one per observation. 
This could lead to a large number of generated tests, each of which is testing the same method with a different input-output pair. 
The Test Selector component produces a test suite from the available test cases. 
Currently, we implement a simple greedy test selection algorithm, 
guided by a single test objective of coverage.
For the single-objective (test coverage) test-selection optimization problem,
a greedy algorithm is known to be efficient, and usually not too far from a  global optima \cite{symh:regression-survey}.

\noindent
{\bf 
Linting:} 
The publisher runs standard Meta CI linters to fix formatting issues and other lint-level issues, such as unused imports. 
We found that it was  easier to fix these using  the existing lint tooling,  as a post-processor, rather than trying to statically fix them at test generation time.
Using the standard linters, which interpose in continuous integration also insures that the diff that is ultimately published, does indeed meet all current linting standards and, therefore, avoids unnecessary spurious lint error messages on the generated diffs.\looseness=-1 

\noindent
{\bf 
Diff Submission:} 
Finally, the publisher submits the generated  diff into the standard Meta CI system. 
This is the mechanism by which TestGen’s generated tests enter production; they go through the diff review process just like any other unit test.

\section{Results}
This section describes our results from deploying TestGen at Meta for Instagram. We report 3 different results: TestGen test results when running in CI at a small scale for 6 months; TestGen coverage at large scale from E2E tests; and back testing TestGen against past large regressions captured in launch blocking tasks (LBs); in the rest of these section. 

\subsection{Running TestGen in Meta CI}
In our deployment process we decided to land TestGen cautiously and incrementally, and with full human code review, rather than automatically landing every test that TestGen could generate (which could potentially run to hundreds of thousands of test cases). 
In this way, we iron out deployment and scaling issues incrementally and avoid bombarding our engineers with a large amount  of unfamiliar test signal all at once. 
While so-doing we also look out for any problems that the deployment of tests at such scale could create for  CI and other parts of Meta infra.
This section answers the following research question:\looseness=-1  

\researchquestion{What is the impact of running TestGen tests in Meta CI on the relevant code changes that Meta engineers submitted for the Instagram Android app?}

In particular, the RQ  establishes baseline statistics for how many tests have landed into Meta production, and how this initial deployment has behaved in production. 
For example, how many times the tests have been executed, and how many bugs they have found so far in production.
\autoref{tab:testgen_ci} presents these  results.
It gives top level statistics on  the TestGen tests that we landed so far and their runs in CI since mid 2023.

The number of blocked diffs is a measure of bug-revealing potential. 
However, it should be noted that these diffs could also have received  signal from other testing and build infrastructure. 
Also, the results include experimental diffs.
That is, engineers often put up initial versions of their changes purely in order to stimulate the testing system to report bugs, fully expecting to see such signal.

Therefore, although these results indicate the bug-revealing potential of the technology, they are not necessarily an accurate reflection of the number of bugs that occur in development, but more reflection of the speed at which developers operate, safe in the knowledge they have a reliable infrastructure that reports bugs. 
As the results show, the TestGen tests do run reliably, and do find bugs. 
This answers RQ1:\looseness=-1

\begin{table}
  \centering
  
  \begin{tabular}{cccc}
    \toprule
     Landed  & Total  & Diff Time  & Blocked  \\
          Tests &  Runs & Failures &  Diffs \\
    \midrule
      518 & 9,617,349 & 75,722 & 5,702 \\
    \bottomrule
    \smallskip
  \end{tabular}
  \caption{Top level TestGen statistics: 518 landed TestGen tests executed a total of 9,617,349 times, failing on 75,722 test executions and detecting 5,702 faults.}
  \label{tab:testgen_ci}
\end{table}

\researchquestionanswer{
TestGen has so-far landed 518 distinct unit tests. 
In the 6 months between July and December 2023, these tests run in Meta CI  a total of $9,617,349$ times. 
TestGen tests failed on $75,722$ of these $9,617,349$ executions, thereby revealing bugs in $5,702$ code changes that, if landed to production,  would have caused regressions.
}

\subsection{TestGen Coverage Achieved by Observations Generated from  E2E Test Executions}
TestGen relies on observing End-to-End (E2E) executions of the app to collect the observations that it needs  in order to be able to auto-generate tests. 
There are many ways in which we can generate such observations, such as executing the app manually, running Sapienz \cite{mhetal:ssbse18-keynote}, and running other automated tools that stimulate app execution.

In the Meta CI, one readily available source of end-to-end executions is the execution of Jest E2E tests\footnote{\url{https://jestjs.io/}}.
This is an interesting source of observations, because it comes from E2E tests.
Generating observations in this way implements a form of test `carving' (see \autoref{sec:related}).
The unit test observations are carved from End-to-End tests, while the unit test assertions are generated from the return values.\looseness=-1

This allows us to investigate a form of `coverage completeness' for the carving process; the degree to which the coverage achieved by the E2E tests is replicated at the unit test level, by carving the unit test observations from the E2E test executions.
Therefore, in RQ2, we investigate the TestGen completeness when generating unit tests for all the code that the Jest E2E runs execute:

\researchquestion{For how many classes that all Jest E2E tests cover, can TestGen produce covering unit tests?}

To answer this question, we run all  Jest E2E tests that are marked as `good' 
(that is, they execute reliably) in CI on the instrumented Instagram Android app. 
We collect observations from these app executions and generate tests from the observations. 
We report the number of Kotlin classes for which TestGen generated tests, as a proportion of the total number of Kotlin classes that Jest E2E tests cover.

Overall, we have a total of 5,284 Jest E2E tests for  the Android Instagram app.
These tests cover a total of 14,621 Kotlin files (this is approximately 40\% of the total number of Kotlin files for the Android app). 
Out of these tests, 4,361 are marked as `good' by CI, while 689 marked as broken, and marked as 234 are flaky\footnote{
These numbers concerning good, broken and flaky, are computed by the Meta CI dynamically and thus change daily. 
They do not necessarily reflect typical numbers for an arbitrary day of execution. 
However, we report them here for completeness, so that the reader can fully understand the results we report, as of the census date when we executed TestGen to collect results for RQ2.}. 

We found that, of all 4,361 Jest E2E tests, TestGen automatically generates tests that covered 86\% of the 14,621 Kotlin files that all the Jest E2E tests cover (5,284 tests).
As a source of  TestGen observations, we used  the 4,361  tests marked as `good' by CI on the day we evaluated RQ2. 
The 86\% coverage result is,  therefore, a lower bound, since some of these files would be covered by currently broken or flaky tests that had previously executed and were thus included in coverage statistics for Jest E2E test execution.

In particular, we believe that the remaining 14\% can be (or already are) covered because:
\begin{itemize}
    \item TestGen only runs `good' Jest E2E tests, while the Jest E2E coverage is reported for all. Only 82\% of Jest E2E tests are `good', so TestGen  coverage is inherently underestimated.
    \item TestGen cannot instrument files that execute only in the main Android Dalvik executable (dex file). Loading all TestGen dependencies  in the  main dex is not currently scalable. 
    \item There are a few remaining (relatively obscure, and infrequently used) Kotlin constructs and compiler corner cases that TestGen does not yet fully support.
\end{itemize}

At this scale we cannot know the precise distributions of these three root causes, the first of which leads to an underestimate, while the second two of which are root causes for lack of coverage.
We believe the biggest cause of the three  is highly likely to be the first of the three, suggesting that the lower bound of 86\% retained coverage is a highly  conservative lower bound.
The answer to the RQ2 is thus:

\researchquestionanswer{When running all 4,361 reliable Jest E2E tests, TestGen produced unit tests for 12,827 files. Thus,  a conservative lower bound is that TestGen automatically covers   86\% of the files that the end-to-end tests cover.}

\subsection{Back testing TestGen Against Past Large Regressions}
TestGen's main purpose is  regression testing. 
In order to validate our hypothesis that TestGen is  effective at preventing  regressions that would otherwise impact developer velocity, we back tested TestGen on previous large regressions that landed in master in October and November 2023.

At Meta, once a diff lands in master, it will eventually be pushed to an internal `alpha' release. 
The alpha release uses a monitoring system that creates launch blocking tasks (LBs), thereby protecting the next stage (`beta' release), and ultimately the deployment of the new version into the App Store. 
Thus, until launch-blocking tasks are fixed, the release is blocked.
This is why we regard such launch-blocking tasks to be high–impact regressions, worthy of study. 
We ask the following research question to assess TestGen's performance at protecting the `alpha' release stage from regressions, and thereby reducing  engineering effort, since it is known that shifting left tends to reduce bug fixing effort \cite{alshahwan:software}.

\researchquestion{What proportion of past high-impact regressions (LB tasks) can TestGen automatically generate a test for that would prevent them?}

\subsubsection{Back testing Method}
For back testing, we identified a pool of LB tasks.
These tasks are typically auto-generated to document detected crashes or bugs that will block Instagram's future app launches. 
Further, because we launch apps at a defined cadence, launch blockers present a real threat to submitting Instagram's latest build to the app store. 
These LB tasks are usually prioritized based on severity and manually triaged and fixed by engineers.
From the pool of launch blocking tasks, a total of 16 tasks, previously manually resolved by engineers with clear code fixes, were randomly selected without consultation with those actively working the TestGen project (Table~\ref{tab:lb_task_detail}).

We only consider LB tasks caused by Kotlin code because TestGen for Java currently remains under development. 
With the tasks identified, we  then generated unit tests covering the impacted files.
With these unit tests we can determine whether the test would fail were the fixes  not to be implemented, by reverting the fix. 
Where such generated tests fail, this indicates that the unit tests could have prevented the launch blocking bug from landing into master.\looseness=-1 

\subsubsection{Back testing Results}

        

\begin{table} 
\begin{threeparttable}[b]
\small
    \centering
    \begin{tabular}{rlcl}
        Tasks & Result & Language & Notes \\
         \hline
        13 & confirmed  & Kotlin & Regression caught successfully\\
        2 & unsupported  & Kotlin &  Anonymous objects unsupported\\
        1 & unsupported  & Kotlin &  Private methods unsupported\\
                 \hline
                     \smallskip
    \end{tabular}
    \caption{Back testing results: TestGen was able to prevent 13 out of 16 app-launch-blocking issues. Of the three missed, two where not unit-testable, while the other was a bug in a private method.}
    \vspace{-2em}
    \label{tab:lb_agg}
     \end{threeparttable}
\end{table}

\autoref{tab:lb_agg} summarizes our results when back testing TestGen on the 16 randomly-selected launch-blocking tasks involving Kotlin files. 
TestGen was able to prevent 13 (81\%) of them. 
Out of the 3 (19\%) for which TestGen was not able to generate a unit test that would have caught the regression:
\begin{itemize}
    \item  2(13\%) were not prevented because  they are not unit-testable (they were part of anonymous methods or objects for which TestGen cannot create a unit test because it cannot call them).
    \item 1 (6\%) was not prevented because the method that caused the crash was declared to be a private method.
\end{itemize} 

For the 2 non-unit-testable LB tasks, 
we could use Testability Transformation \cite{mhetal:tse-flag,cadar:tetra,gong:tetra}
to make them unit-testable.

The other remaining uncovered LB task arose due to a bug in a private method. 
We believe that tests for public methods that call  private ones would to catch them, 
but we had not originally planned to use TestGen to directly test private methods.
We decided to not generate tests for private methods, 
since this would couple the generated tests to implementation details (which is generally a bad practice).
However, if we find regressions are prevalent in private methods, and that these are not trapped by tests generated for the public methods that call the private methods, we may revise this position.
\looseness=-1

The answer to our research question is:

\researchquestionanswer{Out of 16 previous high-impact regressions in Kotlin code, captured by launch blocking tasks (LBs), TestGen was able to successfully generate unit tests that would have prevented 81\% of them from becoming launch blocking (\autoref{tab:lb_agg}).\looseness=-1 }

\begin{table}
    \centering
    \small
    \begin{tabular}{llcc}
         Instagram   & Error & \#TestGen & TestGen\\
           Product  & Type & Tests & Caught?\\
           \hline
          Profile & Illegal State Exception & 2 & Yes \\
         Stories & Uninitialized Property Access & - & No\\
         Stories & Illegal Argument &1& Yes \\
         Deep Links Infra & Class Cast Exception &  1 &  Yes\\
          Comments & Illegal Argument &  1  & Yes\\
         Ads & Null Pointer Exception & - & No  \\
        Direct & Class Cast Exception & 1 &  Yes\\
         OnDevice Tech & Illegal State Exception & 3 & Yes\\
         Feed & Null Pointer Exception & 1 & Yes\\
         Camera & Runtime Exception & 1 &Yes \\
         Camera & Android native crash &  1 &Yes  \\
         Direct & Illegal State Exception & 1 &Yes \\
         Design Systems &  Resources Not Found  &  1  & Yes \\
         Well Being &  Null Pointer Exception &   - & No   \\
         Camera & No Such Method &  1 &Yes \\
         Notifications &  Null Pointer Exception  &   1 &Yes \\
               \hline
               \smallskip
    \end{tabular}
    \caption{Launch blocking tasks on which we back tested TestGen. All these tasks were caused by crashes.}
    \vspace{-2em}
    \label{tab:lb_task_detail}
\end{table}

\section{Related Work}
\label{sec:related}
TestGen is related to Test 
Carving
\cite{elbaum:carving}.
Test carving takes an existing end-to-end test, and carves the unit test from it.
As formulated by Elbaum et al. in 2009 \cite{elbaum:carving}, carving seeks to carve out a `pre-state' from the entire state of the system under test, and a `post state'. 
A regression test is then formed by re-establishing the pre-state and asserting the post state. 
Our approach is similar to this, in the sense that the pre-state, for our observation-based tests, consists of the values of the parameters observations of the method under test, together with  the current object instance, while the post state is the observation of return values.

TestGen generates unit level tests.
As shown by Gross, Fraser and Zeller in 2012 \cite{gross:false},   unit test generators may generate false positives because the unit level parameters and return values are not realizable in any system-level execution. 
They reported that all 181 errors found using a popular random test generation system on five open source systems were false positives. 
TestGen circumvents this potential false positive  issue using an observation-based approach with tests generated from observations of {\em real} end-to-end executions.\looseness=-1

Kampmann and Zeller \cite{kampmann:carving}, overcome the false positive issue by carving {\em parameterized} unit tests, such that the resulting unit test can be `lifted' back to a system level test. 
This allows possible false positive  test failures to be detected, by checking whether the lifted computation raises a system failure or not (unit test failures that do not lift to  system test failures are discarded as false positives).  
Kampmann and Zeller implemented this for C, and showed how it allowed them to extend carving to make it applicable for fuzzing~\cite{manes:fuzzing}.\looseness=-1 
Other authors extended the notion of carving, most recently focusing on carving UI tests to generate API tests \cite{yandrapally:carving}, where the goal is to dynamically infer suitable API endpoints, 
so that UI-based testing can be carved to stimulate the corresponding API endpoints.

TestGen's use of serialized observations is related to previous work on object capture, such as the OCAT system, which is used to capture objects for automated testing \cite{jaygarl:ocat}. 
Jaygarl et al. applied OCAT to three open source systems, after integrating it with Randoop \cite{pacheco:randoop}. 
They also mutate the captured observations, although there is no mechanism for avoiding the consequent potential for false positives that this may introduce.

Finally, TestGen requires sophisticated serialization and deserialization.
It is, therefore, also related to previous work on serialization and deserialization, such as XStream\footnote{https://x-stream.github.io/} and JSX\footnote{https://facebook.github.io/jsx/}.
One of the primary contributions of the present paper is the introduction of a scalable, depth-aware, fully-featured serialization and deserialization framework, suitable for capturing the complex object serializations needed by industrial observation-based testing.

One of the other potential applications of observation-based testing, is the use of deserialized objects in human authoring of test cases. 
As reported recently by Fazzini et al. \cite{fazzini:use} developers typically go to some considerable length to mock, partly spy, fake, and otherwise build complex objects in order to execute unit tests. 
The observations from observation-based test generation can be used to circumvent the need for the developer to spend (what is typically  considerable) time and effort building such `test double' objects.\looseness=-1

The idea of testing based on observations has been in the literature for over two decades, dating back to the JRapture system for Java, introduced in 2000 \cite{steven:jrapture}.
Both TestGen and previous observation-based capture/replay tools share the same orientation towards regression testing, based on captured observations of runtime values.  
However, while TestGen's observations are serialized high level object instances captured from the instrumentation of source code, JRapture captures low level interactions at bytecode level.
Other more recent observation-based testing approaches have also used higher-level observations captured, for example, for automotive systems \cite{wolschke:observation}.
However, in this automotive work, the observations are of system-level scenarios, mined from traces of simulations and real behaviors for replay in simulation.
That makes these automotive system approaches more like Meta's simulation-based testing approach \cite{jaetal:ease21-keynote}, than its TestGen observation-based approach, for which observations  are serialized object values.

TestGen is also related to existing work on generation of tests more generally.
There have been a great many techniques proposed for test generation over the past five decades, starting with the early work on random test selection in the mid-1970s \cite{boyer:select}.
An account of the many and varied techniques for test data generation would require a dedicated paper in its own right. 
For brevity, we refer the reader to existing comprehensive surveys \cite{anandetal:orchestrated,cadar:three-decades,mcminn:survey}.

TestGen's primary novelty and technical advance, lies in the way it tackles the challenges inherent when collecting  observations at scale.
Without this, TestGen could not achieve an observation-based test generation technique that operates on apps like Instagram, that consist of many tens of millions of lines of code.

\section{Future Work}
\label{sec:future}

TestGen  is designed to allow easy incorporation of other test generation strategies.
It will be particularly interesting to explore hybrid approaches that extend existing approaches to test generation, such as 
concolic \cite{cadar:three-decades},
fuzzing-based \cite{manes:fuzzing},
random \cite{pacheco:randoop} and 
search-based \cite{mh:icst15-keynote}, hybridizing each to additionally incorporate observations.

Future work may also consider observation-based approaches to Metamorphic Testing \cite{chen:survey,segura:metamorphic-survey}. 
While TestGen currently targets regression testing, there is an established link between regression testing and Metamorphic Testing \cite{jaetal:mia}, which has also recently been deployed at scale in industry at Meta \cite{jaetal:mia,kbmh:met22-keynote} and at Google \cite{donaldson:metamorphic}.
In future work, it may be possible to adapt existing metamorphic relation inference algorithms \cite{su:dynamic,zhang:search-based-MR} to apply them to observations. 
This would allow us to extend TestGen 
to tackle metamorphic testing, especially where such algorithms already have available implementations that target Java/Kotlin documentation \cite{blasi:memo}, thereby residing within the same programming language domain as TestGen. 

We would be excited to partner with the academic community on collaborative projects to extend these techniques to consider observation-based formulations,  which we can then collaboratively  evaluate at scale, using the TestGen framework in place at Meta.

TestGen could  hybridize with other observation-based approaches, such as 
Observation-based Slicing \cite{dbetal:orbs-fse14}.
Observation-based Slicing may also be helpful to TestGen, since it may help identify suitable boundaries for the introduction of mocks. 
TestGen does not currently generate mocks, but fortunately the core problem has previously been studied
\cite{alshahwan:automock,saff:mock} and extensions that do create mocks  could be informed by work on decisions about when to mock \cite{spadini:mock}. 
Observation-based Slicing, combined with these existing mocking techniques, may additionally provide powerful observation-based dependence analysis, flexible mocking and debugging.\looseness=-1

Future work may also consider multi-objective regression test formulations \cite{mh:morto-keynote} to balance different competing regression test criteria, and may therefore extend test selection to use other selection algorithms.
Future work might also draw on invariant inference, such as techniques like Daikon \cite{ernst:dynamically-tse}
to infer likely invariants, over observations, and use these to detect abnormal behaviors, thereby extending TestGen from regression testing to other forms of testing.


\section{Conclusion}
This paper introduced TestGen, a scalable observation-based unit test carver, developed by Meta’s Instagram Product Foundation team, since October 2022, and first deployed in July 2023. So far,  TestGen has reported  5,702 bugs, while back testing demonstrated it can trap at least 81\% of high-impact regressions that would otherwise become app launch blocking for Instagram.

\bigskip

\noindent {\bf Acknowledgements:}
\small
We wish to thank the  Instagram organization and leadership for their support, and the many Meta engineers who helped us to deploy and evaluate TestGen. 
\normalsize

\newpage
\balance

%
\bibliographystyle{ACM-Reference-Format}
\bibliography{main}

\end{document}